\documentclass[12pt]{article}
\usepackage{pdproc, epsfig} 
\usepackage{amssymb}
\usepackage{url}
\usepackage[dvips]{hyperref}
\def\be{\begin{equation}}
\def\ee{\end{equation}}
\def\bea{\begin{eqnarray}}
\def\eea{\end{eqnarray}}
\def\nn{\nonumber}
\begin{document}
\renewcommand{\thefootnote}{\alph{footnote}}

\title{
A MONOCHROMATIC NEUTRINO BEAM FOR $\theta_{13}$ and $\delta$\footnote{Proc.~3rd.~International Workshop on NO-VE: Neutrino Oscillations in Venice, 7-10 February 2006, Venice.}}

\author{JOSE BERNABEU and CATALINA ESPINOZA}
\address{Universitat de Val\`encia and IFIC,\\ 
   E-46100 Burjassot, Val\`encia, Spain\\ 
 {\rm E-mail: Jose.Bernabeu@uv.es}\\
 {\rm E-mail: m.catalina.espinoza@uv.es}}

\abstract{The goal for future neutrino facilities is the determination  of the $[U_{e3}]$ mixing and CP violation in neutrino oscillations. This will require precision experiments with a very intense 
neutrino source. Here a novel method to create a monochromatic neutrino beam based on the recent discovery of nuclei that decay fast through 
electron capture is discussed. The boost of such radioactive ions  will generate a monochromatic directional neutrino beam when decaying 
at high energy in a storage ring with long straight sections. We show that the capacity of such a facility to discover new physics is impressive, so that the principle of energy dependence in the oscillation probability of the $\nu_e \to \nu_{\mu}$ channel is operational to separate out the two parameters of the mixing  $\theta_{13}$ and of the  CP-violating phase $\delta$.}
   
\normalsize\baselineskip=15pt

\section{Introduction}

\noindent Neutrinos are very elusive particles that are difficult to detect. Even so, physicists have
over the last decades successfully studied neutrinos from a wide variety of sources, either natural,
such as the sun and cosmic objects, or manmade, such as nuclear power plants or accelerated beams.
 Spectacular results have been obtained in the last few years for the flavour mixing of neutrinos
 obtained from atmospheric, solar, reactor and accelerator sources and interpreted in terms of the
 survival probabilities for the beautiful quantum phenomenon of neutrino oscillations
\cite{fukuda,ahmad}. The weak interaction eigenstates  $\nu_{\alpha}$ $(\alpha = e,\mu, \tau)$
 are written in terms of mass eigenstates
$\nu_k$ $(k=1,2,3)$ as  $\nu_{\alpha} = \sum_k U_{\alpha k} (\theta_{12}, \theta_{23},
\theta_{13};\delta ) \nu_k$, where $\theta_{ij}$ are the mixing angles among the three neutrino
families and $\delta$  is the $CP$ violating phase. Neutrino mass differences and the mixings for
the atmospheric $\theta_{23}$ and solar $\theta_{12}$ sectors have thus been determined. The third connecting mixing $\vert U_{e3} \vert$ is bounded as  $\theta_{13}\le 10^{\circ}$   from the CHOOZ
reactor experiment \cite{apollonio}. In Section~2 we present what we do know on the properties of massive neutrinos as well as what is still unknown and searched for in ongoing and future experiments. General considerations about  the discrete symmetries are given in Section~3, whereas Section~4 discusses the virtues of the suppressed oscillation channel $\nu_e \to \nu_{\mu}$. 
Next experiments able to measure the still undetermined
mixing  $\vert U_{e3} \vert$ and the $CP$ violating phase $\delta$, responsible for the matter-antimatter asymmetry, need
to enter into a high precision era with new machine facilities and very massive detectors. As neutrino oscillations are energy dependent, for a given baseline, we consider  a facility able to study the detailed energy dependence by means of fine tuning of boosted monochromatic
neutrino beams from electron capture \cite{Bernabeu:2005jh}. In such a facility, the neutrino energy is dictated by the chosen boost
of the ion source and the neutrino beam luminosity is concentrated at a single
known energy which may be chosen at
will for the values in which the sensitivity for the $(\theta_{13}, \delta)$ parameters is higher. The analyses showed that this concept could become operational only
 when combined with the recent discovery of nuclei far from the stability line, having super
 allowed spin-isospin transitions to a giant Gamow-Teller resonance kinematically accessible
 \cite{algora}. In Section~5 we will develop the monochromatic neutrino beam concept and we give details about its implementation using such short-lived ions.

In Section~6, the Neutrino Flux emerging from the facility
with boosted decaying ions is calculated and the main characteristics discussed. In Section~7, we show the sensitivity which can be reached with the proposed facility for the parameters
 $(\theta_{13}, \delta)$ of neutrino oscillations. Some conclusions and outlook are given in
Section~8.

\section{Status of Neutrino Properties}\label{sec2}
\noindent The most sensitive method to prove that neutrinos are massive is provided by neutrino oscillations \cite{pontecorvo}. These phenomena are quantum mechanical processes based on masses and mixing of neutrinos. The fundamental statement is that the weak interaction states (greek indices) do not coincide with the mass eigenstates (latin indices) and are rather given by the coherent superposition
\be\label{superposition}
\nu_{\alpha}=\sum_k U_{\alpha k}\nu_k,
\ee
where $\nu_k$ can be either Dirac or Majorana particles. For flavour oscillations, the general mixing for three families is parametrized by three angles and one $CP$-phase, accompanying two independent mass differences $(\Delta m_{ij}^2=m_i^2-m_j^2)$. The usual factorization of the mixing matrix $U$ is given by
\bea
U =
\left[\begin{array}{ccc}
1  & 0  & 0 \\
0  & c_{23} & s_{23}  \\
0  & -s_{23} & c_{23}
\end{array}
\right]
\left[\begin{array}{ccc}
 c_{13} & 0  & s_{13}\, e^{-i\delta} \\
0  & 1 & 0  \\
-s_{13}\, e^{i\delta}  & 0 & c_{13}
\end{array}
\right]
\left[\begin{array}{ccc}
 c_{12} & s_{12} & 0 \\
-s_{12}  & c_{12} & 0  \\
0  & 0 & 1
\end{array}
\right]\,\,\,\,\,,
\label{Upartes}
\eea
where $c_{ij}=\cos{\theta_{ij}}$ and $s_{ij}=\sin{\theta_{ij}}$.  This parametrization is an interesting form to cast the mixing matrix because it separates the contributions comming from atmospheric and solar neutrinos. The left matrix is probed by atmospheric neutrinos and long-baseline neutrino beams, when looking  for either $\nu_{\mu}$-disappearance or $\nu_{\mu} \to \nu_{\tau}$ appearance, the right matrix by solar neutrinos and long-baseline reactor experiments. The main question at present is the search of appropriate experiments to probe the middle connecting matrix which contains fundamental information about the mixing $\theta_{13}$ and, when combined with the others, on $CP$ violating phenomena.

If neutrinos are Majorana, the mixing matrix  incorporates two additional physical phases that can only become apparent in processes with a Majorana neutrino propagation, violating global lepton number in two units, $(\Delta L)=2$. As long as one looks for flavour oscillations, $U$ describes the mixing even if neutrinos are Majorana particles.

At present, there are several pieces of evidence for neutrino oscillations. The results of solar neutrino experiments (Homestake \cite{Cleveland:1998nv}, Kamiokande \cite{fukuda2}, SAGE \cite{Abdurashitov:2002nt}, GALLEX \cite{gallex}, GNO \cite{Altmann:2005ix}, Super-Kamiokande \cite{Hosaka:2005um} and SNO \cite{ahmad,Aharmim:2005gt}) and the reactor long-baseline experiment KamLAND \cite{Araki:2004mb} have measured $\sin^22\theta_{12}\sim 0.81$ and the square mass difference $\Delta m_{12}^2=8\times 10^{-5}\; eV^2$. Atmospheric neutrino experiments (Kamiokande \cite{Kajita:1998bw}, IMB \cite{imb}, Super-Kamiokande \cite{fukuda,Ashie:2005ik}, Soudan-2 \cite{Allison:2005dt} and MACRO \cite{Ambrosio:2003yz}) and the accelerator K2K experiment \cite{Aliu:2004sq}, together with the negative results of the CHOOZ experiment \cite{Apollonio:2002gd}, have constrained $\sin^22\theta_{23}=1.00$ and $\Delta m_{23}^2=2.4\times 10^{-3}\; eV^2$. The CHOOZ reactor experiment places an upper bound for the third connecting mixing, $\theta_{13}\le 10^{\circ}$  \cite{apollonio}.

One should realize that Eq.(\ref{superposition}) with only three light active neutrinos is incompatible with LSND result \cite{athana}. One would need at least one additional sterile neutrino mixed with active neutrinos. MiniBoone experiment will settle this question \cite{miniboone}.

Neutrino oscillation experiments are not able to measure absolute neutrino masses but only differences of masses-squared. To fix the absolute mass scale, direct neutrino mass searches like beta decay and double beta decay are needed.

\begin{figure}[ht!]
\centering
\includegraphics[width=6cm,height=5cm]{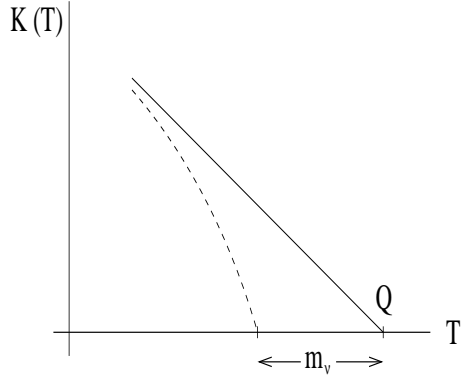}
\caption{Kurie plot for $^3H$ beta decay.}
\label{betadecay}
\end{figure}

Fermi proposed \cite{fermi} a kinematic search of neutrino mass from the hard part of the beta
spectra in $^{3}H$ beta decay. The ``classical'' decay $^{3}H\rightarrow ^{3}He+e^{-}+\overline{\nu }_{e}$ is a superallowed transition with a very small energy release $Q=18.6$ $KeV$. As it can be seen in the Kurie plot (see Fig.\ref{betadecay}), a non-vanishing neutrino mass $m_{\nu}$ provokes a distorsion from the straight-line $T$-dependence at the end point of the energy spectrum, $T$ being the kinetic energy of the released electron. As a consequence, $m_{\nu }=0\rightarrow T_{\max }=Q$ whereas $m_{\nu }\neq 0\rightarrow T_{\max }=Q-m_{\nu}$.

The most precise Troitsk and Mainz experiments \cite{mainz,weinheimer} give no indication in
favour of $m_{\nu }\neq 0$. One has the upper limit $m_{\nu }<2.2$ $eV$ $(95\%$ $CL$). In the near future, the KATRIN experiment \cite{katrin} will reach a sensitivity of about $0.2$ $eV$. In fact, if the energy resolution were $\Delta T<<m_{\nu}$, one would see three different channels for $\beta$-decay, one for each mass-eigenstate neutrino. At present, with $\Delta T >> m_{\nu}$, one sees an incoherent sum \cite{Farzan:2002zq} $m_{\nu}^2=\sum_j|U_{ej}|^2m_j^2$ of the three channels.

\begin{figure}[ht!]
\centering
\includegraphics[width=4cm,height=5cm]{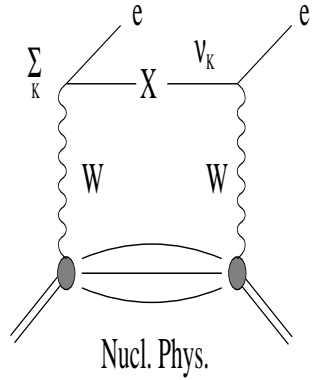}
\caption{Neutrinoless double-$\beta$ decay.}
\label{doblebetadecay}
\end{figure}

Still we do not know whether neutrinos are Dirac or Majorana particles. Neutrinoless double-$\beta$ decay is a very important process, mainly because it is the best known way to distinguish Dirac from Majorana neutrinos
\cite{furry}. Neutrinoless double-$\beta$ decays are processes of type $(A,Z)\to (A,Z+2)+e^{-}+e^{-}$. They are allowed for Majorana neutrino virtual propagation. In Fig.\ref{doblebetadecay} it is represented as a second order weak interaction amplitude

\noindent where $-X-$ indicates a Majorana neutrino mass insertion $\nu_L^C \to \nu_L$ with $\Delta L =2$. The expression of the neutrinoless probability is factorized in different ingredients
\be\label{doblebetaamp}
\mathrm{Prob}[\beta\beta_{0\nu}]=(\mathrm{Phase\,Space})|<m_{\nu}>(\mathrm{Nuclear\, Physics})|^2.
\ee

The quantity of primary interest in neutrino physics is the average neutrino mass $<m_{\nu}>=\sum_k U_{ek}^2 m_k$, where $U_{ek}^2$ is for Majorana neutrinos. Notice the sensitivity to the phases of $U$ and not only to moduli. This result shows that the main ingredient to produce an allowed $(\beta \beta)_{0 \nu}$ is the massive Majorana neutrino character. The expression (\ref{doblebetaamp}) shows the dependence of the probability with the absolute neutrino masses, not with the mass differences. Under favourable circumstances, a positive signal of the $(\beta \beta)_{0 \nu}$ process could be combined with results of neutrino oscillation studies to determine the absolute scale of neutrino masses \cite{bilenky}.
A possible indication of $(\beta \beta)_{0 \nu}$ for $^{76}Ge$ has been discussed in 
\cite{Klapdor-Kleingrothaus:2004wj}. But the experimental status is uncertain, taking into account the limits given by the Heidelberg-Moscow \cite{Klapdor-Kleingrothaus:2001yx} and IGEX \cite{Aalseth:2002rf} collaborations.

\begin{figure}[ht!]
\centering
\includegraphics[width=8cm,height=8cm]{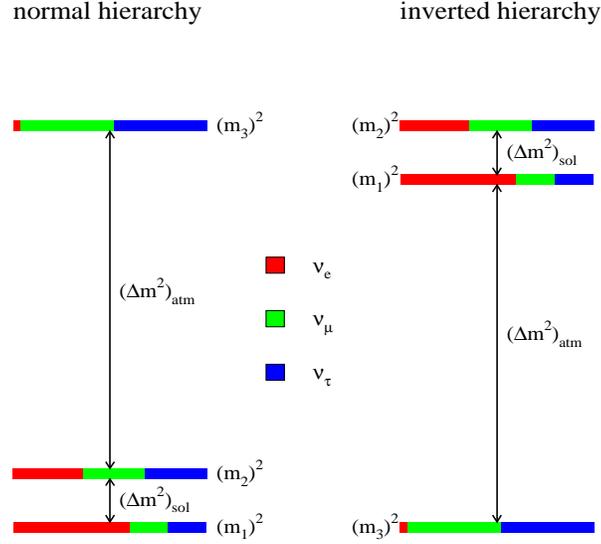}
\caption{The two possible neutrino-mass herarchies.One shows as different shadows the flavour content of each mass-eigenstate.}
\label{herarchy}
\end{figure}

One question which cannot be settled by neutrino oscillation in vacuum is the form of the spectrum for massive neutrinos, either hierarchical or inverted, as shown in Fig.\ref{herarchy}. This is because the vacuum neutrino oscillations depend just on the square of the sine of mass differences and are therefore independent of the sign. But the interference with medium effects could see the sign of $\Delta m^2$.

In 1985 Mikheev and Smirnov \cite{smirnov}, building on the earlier work of Wolfenstein \cite{wolfenstein}, realized that interactions of the neutrinos with matter in the Sun or even the Earth could lead to a substantial  modification of the oscillations (now called the MSW effect). When propagating through matter the free-particle Hamiltonian must be modified to include the charged current forward elastic scattering amplitude of electron neutrinos with electrons, the only piece which builds a different phase for the three neutrino species.  A similar analysis proposed recently for atmospheric neutrinos 
\cite{Banuls:2001zn} opens the way  for matter effects measurements sensitive to the sign of $\Delta m^2_{13}$. After diagonalization of the Hamiltonian, the result of such analysis for the effective mixing $\tilde{\theta_{13}}$ is given by
\be
\sin^22\tilde{\theta_{13}}=\frac{\sin^22\theta_{13}\left(\frac{\Delta m_{13}^2}{a}\right)^2}{\left(E-\cos2\theta_{13}\frac{\Delta m_{13}^2}{a}\right)^2+\sin^22\theta_{13}\left(\frac{\Delta m_{13}^2}{a}\right)^2}
\ee
where $a=\sqrt{2}G_FN_e$. $N_e$ is the electron number density in matter, $G_F$ is the Fermi coupling constant, and $E$ is the energy of the neutrino. The last equation shows the possibility of a resonant MSW behaviour at a energy $E_R=\cos 2\theta_{13}\Delta m_{13}^2/a$.
 In going from $\nu$ to $\bar{\nu}$, the matter-term changes sign $a \to -a$, so that the MSW resonance will be apparent either for neutrinos or for antineutrinos.
For small $\theta_{13}$, the resonance could provide a clean measure of the sign of $\Delta m_{13}^2$. Indeed, for $\Delta m_{13}^2>0$ the resonance appears only for neutrinos, whereas for $\Delta m_{13}^2<0$ it would show up only for antineutrinos. This effect can be observed either with a magnetized iron detector, able to have charge discrimination, or with a water Cherenkov detector using \cite{Bernabeu:2003yp} the different cross section for neutrinos and antineutrinos.

The data from Super-K, SNO, K2K, KamLAND have established a solid evidence of neutrino oscillations. New measurements at Super-K, SNO, KamLAND, K2K, Borexino, Minos, CNGS should improve our knowledge of the atmospheric and solar parameters. But there is still much work to be done in future facilities.  One of the main pending questions in the determination of the mixing matrix $U$ concerns the $\theta_{13}$ ingredient closely related to the CP-violating phase $\delta$. The value of $\theta_{13}$ is going to be searched for in the accelerator T2K experiment \cite{t2k} and the reactor DOUBLE CHOOZ collaboration \cite{Lasserre:2004vt}. The problem of CP-violation in the lepton sector awaits a decision on new proposed facilities such as super beams, beta beams or neutrino factories. In the following we develop a novel proposal \cite{Bernabeu:2005jh} aimed to shed light on those questions of $\theta_{13}$ and $\delta$.

\section{Symmetries in Neutrino Oscillations (CP, T, CPT)}\label{sec3}
\noindent Complex neutrino mixing for three neutrino families originates CP and T violation in neutrino oscillations. The requirement of CPT invariance leads to the condition~\cite{Bernabeu:1999gr,Bernabeu:1999ct}
$A\left( \overline{\alpha} \rightarrow \overline{\beta};\ t \right) 
= A^* \left( \alpha \rightarrow \beta;\ -t \right)$ for
the probability amplitude between flavour states ($A(\alpha \to \beta;t)=\sum_{k}U_{\alpha k}U^{*}_{\beta k}\exp[-iE_kt]$), so that CP or T violation
effects can take place in appearance experiments only.

CPT invariance and the unitarity of the mixing matrix imply that the
CP-odd probability
\be
D_{\alpha \beta}=\left|  A\left(  \alpha\rightarrow\beta; \ t\right)  \right|^2
-\left| A\left ( \overline{\alpha} \rightarrow \overline{\beta};\ t\right ) \right |^2
\label{eq:cpodd}
\ee
is unique for three flavours: $D_{e \mu}=D_{\mu \tau}=D_{\tau e}$. There are, however, three independent CP-even probabilities 

\be
S_{\alpha \beta}=\left|  A\left(  \alpha\rightarrow\beta; \ t\right)  \right|^2
+\left| A\left ( \overline{\alpha} \rightarrow \overline{\beta};\ t\right ) \right |^2
\label{eq:indepprob},
\ee
as given by $S_{e \mu}$, $S_{\mu \tau}$ and $S_{\tau e}$.
The T-odd probabilities
\bea
T_{\alpha \beta}&=&\left| A\left(\alpha \rightarrow \beta;\ t\right) \right|^2
-\left| A\left(\beta \rightarrow \alpha;\ t\right) \right|^2, \nn \\
\overline{T}_{\alpha \beta}&=&
\left|A\left( \overline{\alpha}\rightarrow \overline{\beta};\ t\right)\right|^2
-\left|A\left(\overline{\beta}\rightarrow\overline{\alpha};\ t\right)\right|^2
\label{eq:todd}
\eea
are odd functions of time~\cite{Bernabeu:1999gr,Bernabeu:1999ct} by virtue of the hermitian
character of the evolution Hamiltonian. 
The last property does not apply to the effective Hamiltonian of the 
$K^0 -\overline{K}^0$ system.

The oscillation terms are controlled by the phases 
$\Delta_{ij} \equiv \Delta m_{ij}^{2} L/4E$,
where $L=t$ is the distance between source and detector. This suggests the performance of energy dependent measurements to separate out the different parameters.
In order to generate a non-vanishing CP-odd
probability, the three families have to participate actively: in the limit
$\Delta_{12} \ll 1$, which refers to the lowest difference of mass eigenvalues, 
the effect tends to vanish linearly with 
$\Delta_{12}$. 
In addition, all mixings and the CP-phase have to be non-vanishing. 
If $\Delta_{12} \ll 1$, the flavour transitions can be classified
by the mixings leading to a contribution from the main
oscillatory phase $\Delta_{23} \simeq \Delta_{13}$.
The strategy would be the selection of a 
``forbidden'' transition, i.e., an appearance channel with very low
CP-conserving probability, in order to enhance the CP-asymmetry. 
This scenario appears realistic from present results on neutrino
masses and mixings from atmospheric and solar neutrino experiments, with the selection of the $\nu_e \to \nu_{\mu}$ appearance transition. 

\section{Suppressed $\nu_e \to \nu_{\mu}$ Oscillation}\label{sec4}
\noindent The observation of $CP$ violation needs an experiment in which the emergence of another neutrino flavour
 is detected rather than the deficiency of the original flavour of the neutrinos. At the same time, the interference needed to generate CP-violating observables can be enhanced if both the atmospheric and solar components have a similar magnitude. This happens in the suppressed $\nu_e \to \nu_{\mu}$ transition. The appearance
 probability $P(\nu_e \to \nu_{\mu})$ as a function of the distance between source and detector
$(L)$ is given by \cite{cervera}

\bea\label{prob}
P({\nu_e \rightarrow \nu_\mu}) & \simeq &
s_{23}^2 \, \sin^2 2 \theta_{13} \, \sin^2 \left ( \frac{\Delta m^2_{13} \, L}{4E} \right ) +
c_{23}^2 \, \sin^2 2 \theta_{12} \, \sin^2 \left( \frac{ \Delta m^2_{12} \, L}{4E} \right )
\nonumber \\
& + & \tilde J \, \cos \left ( \delta - \frac{ \Delta m^2_{13} \, L}{4E} \right ) \;
\frac{ \Delta m^2_{12} \, L}{4E} \sin \left ( \frac{  \Delta m^2_{13} \, L}{4E} \right ) \, ,
\eea
where $\tilde J \equiv c_{13} \, \sin 2 \theta_{12} \sin 2 \theta_{23} \sin 2 \theta_{13}$.       
   The three terms of Eq. (\ref{prob}) correspond, respectively, to contributions from the
atmospheric and solar sectors and their interference. As seen, the $CP$ violating contribution
has to include all mixings and neutrino mass differences to become observable. The four measured parameters $(\Delta m_{12}^2,\theta_{12})$  and  $(\Delta m_{23}^2,\theta_{23})$ have been fixed throughout this paper to their mean values \cite{Gonzalez-Garcia:2004jd}. 

\begin{figure}[ht!]
\centering
\includegraphics[width=14cm,height=11cm]{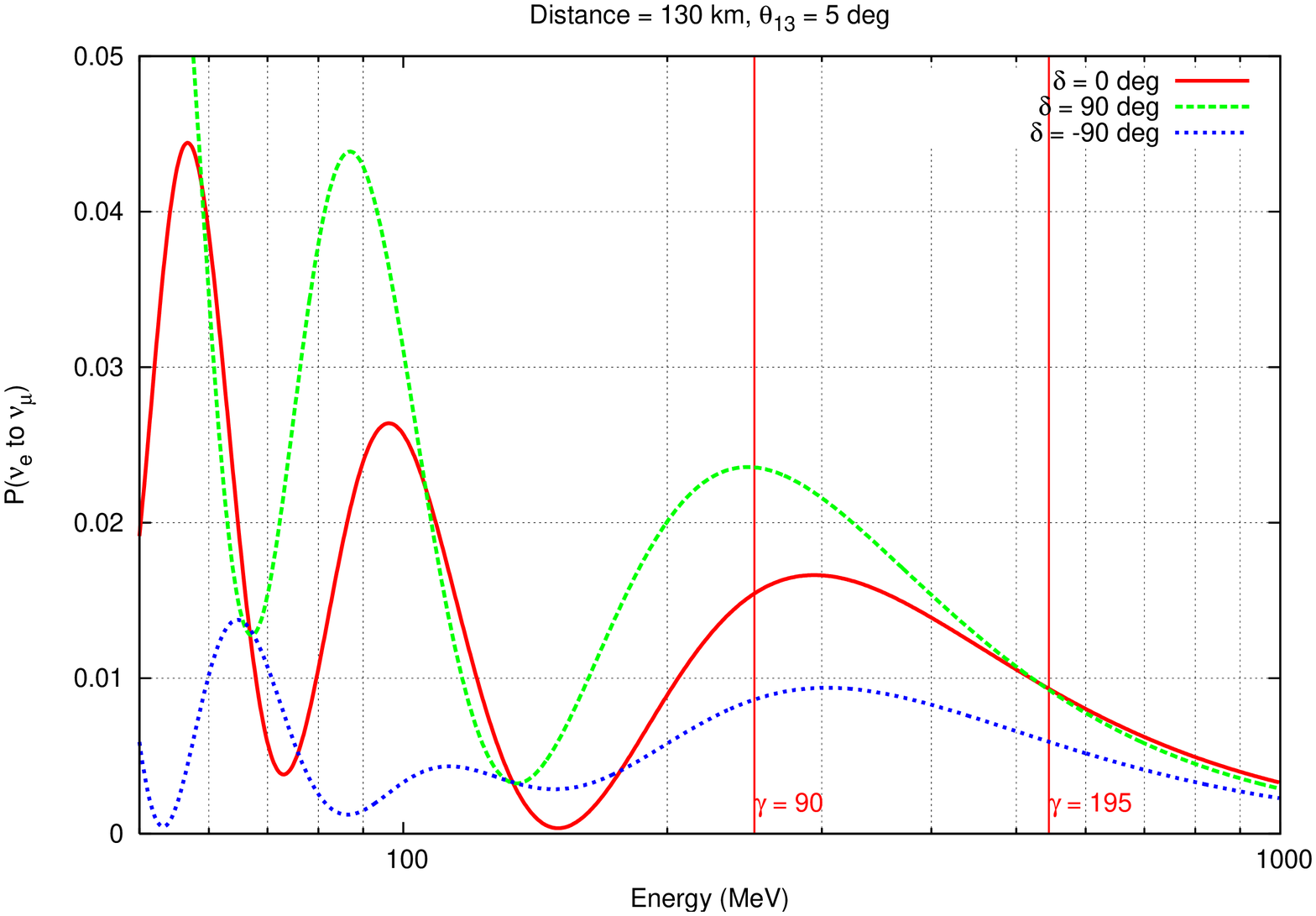}
\caption{The appearance probability $P(\nu_e \rightarrow \nu_{\mu})$ for neutrino oscillations
 as a function of the LAB energy E, with fixed distance between source and detector and connecting
mixing. The three curves refer to different values of the CP violating phase $\delta$. The two
vertical lines are the energies of our simulation study.}
\label{proba}
\end{figure}

Neutrino oscillation phenomena are energy dependent (see Fig.\ref{proba}) for a fixed distance between source and detector, and the observation of this energy dependence would disentangle the two important parameters: whereas $\vert U_{e3} \vert$ gives the strength of the appearance probability, the $CP$ phase acts as a phase-shift in the interference pattern. These properties suggest the consideration of a facility able to study the detailed energy dependence by means of fine tuning of a boosted monochromatic neutrino beam. As shown below, in an electron capture facility the neutrino energy is dictated by the chosen boost
of the ion source and the neutrino beam luminosity is concentrated at a single
known energy which may be chosen at
will for the values in which the sensitivity for the $(\theta_{13}, \delta)$ parameters is higher. This is in contrast to beams with a continuous spectrum, where the intensity is shared between sensitive
and non sensitive regions. Furthermore, the definite energy would help in the control of both the systematics
and the detector background. In the beams with a continuous spectrum, the neutrino energy has to be reconstructed in the detector. In water-Cerenkov detectors, this reconstruction is made from supposed quasielastic events by measuring both the energy and direction of the charged lepton. This procedure suffers from non-quasielastic background, from kinematic deviations due to the nuclear Fermi momentum and from dynamical suppression due to exclusion effects \cite{Bernabeu72}.

In the CERN Joint Meeting of BENE/ECFA for Future Neutrino Facilities in Europe,
the option of a monochromatic neutrino beam from atomic electron capture in $^{150}Dy$ was
considered and discussed both \cite{bernabeu} in its Physics Reach and the machine feasibility.
 This idea was conceived earlier \cite{lindroos0} by the authors and presented together with the
 beta beam facility. The analyses showed that this concept could become operational only
 when combined with the recent discovery of nuclei far from the stability line, having super
 allowed spin-isospin transitions to a giant Gamow-Teller resonance kinematically accessible
 \cite{algora}. Thus the rare-earth nuclei above $^{146}Gd$ have a small enough half-life for
 electron capture processes. Some preliminary results for the physics reach were presented in
 \cite{jordi}. A subsequent paper \cite{Sato:2005ma} appeared in the literature with the proposal of an EC-beam
 with fully stripped long-lived ions. This option would oblige recombination of electrons with ions
 in the high energy storage ring. Such a process has a low cross section and would lead to low
 intensities at the decay point. Even if the production rate would be considerably higher for
 these long-lived nuclei it would result in extremely high currents in the decay ring, something which
 already in the present beta-beam proposal is a problem due to space charge limitations and
 intra-beam scattering. We discuss the option of short-lived ions \cite{Bernabeu:2005jh}.

\section{Electron Capture Process}
\begin{figure}
\centering
\includegraphics[width=11cm,height=10cm]{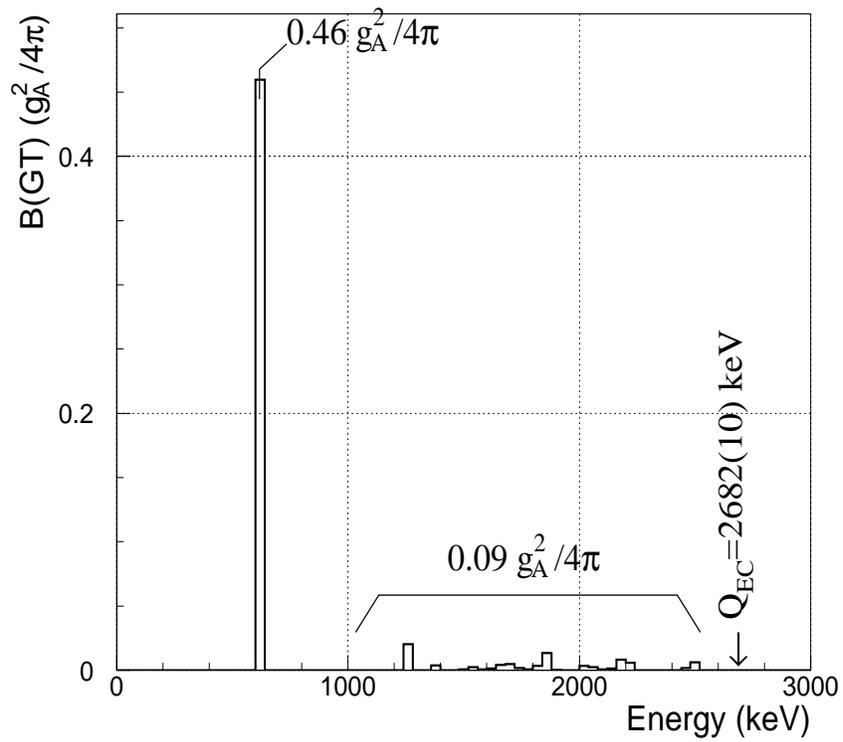}
\caption{Gamow-Teller strength distribution in the $EC/\beta^+$ decay of $^{148}Dy$.}\label{dy}
\end{figure}

\begin{figure}
\centering
\includegraphics[width=14cm,height=8.5cm]{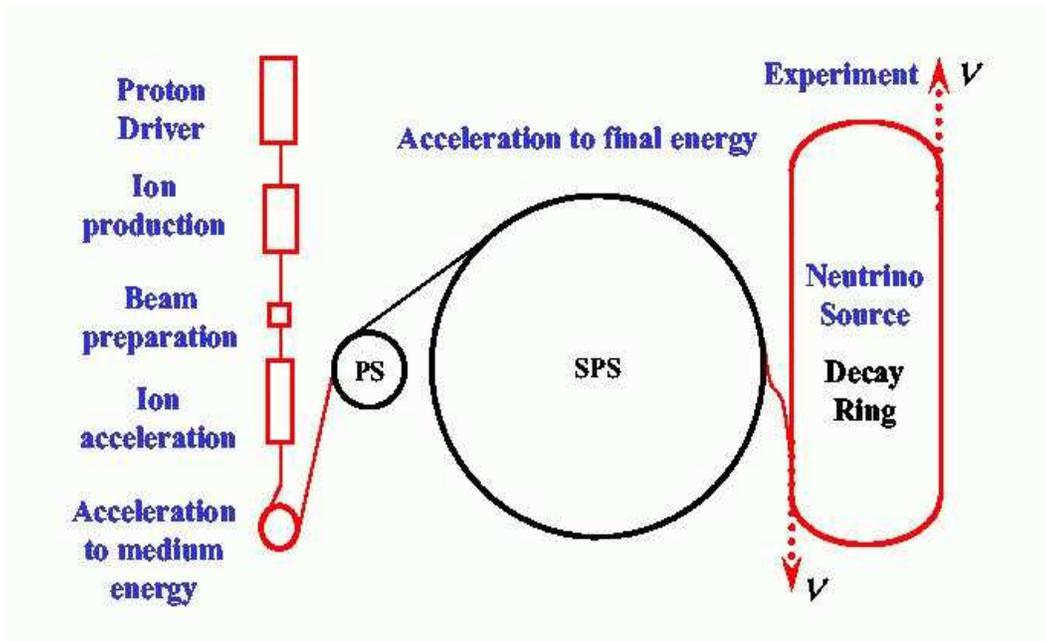}
\caption{A proposal for a CERN EC neutrino beam facility.}
\label{betabeam}
\end{figure}

\noindent Electron Capture is the process in which an atomic electron is captured by a proton
of the nucleus leading to a nuclear state of the same mass number $A$, replacing the 
proton by a neutron, and a neutrino. Its probability amplitude is proportional to the atomic 
wavefunction at the origin, so that it becomes competitive with the nuclear $\beta^+$ decay 
at high $Z$. Kinematically, it is a two body decay of the atomic ion into a nucleus and the 
neutrino, so that the neutrino energy is well defined and given by the difference between
the initial and final nuclear mass energies $(Q_{EC})$ minus the excitation energy of the 
final nuclear state. In general, the high proton number $Z$ nuclear beta-plus decay $(\beta^+)$
 and electron-capture $(EC)$ transitions are very "forbidden", i.e., disfavoured, because the
energetic window open $Q_{\beta}/Q_{EC}$ does not contain the important Gamow-Teller strength
excitation seen in (p,n) reactions. There are a few cases, however, where the Gamow-Teller 
resonance can be populated (see Fig.\ref{dy}) having the occasion of a direct study of the "missing"
 strength. For the rare-earth nuclei above $^{146}Gd$, the filling of the intruder level $h_{11/2}$ 
for protons opens the possibility of a spin-isospin transition to the allowed level $h_{9/2}$
for neutrons, leading to a fast decay. The properties of a few examples \cite{nacher} of
interest for neutrino beam studies are given in Table 1. A proposal for an accelerator facility
with an EC neutrino beam is shown in Fig.\ref{betabeam}. It is based on the most attractive features of the
 beta beam concept \cite{zucchelli}: the integration of the CERN accelerator complex and the
 synergy between particle physics and nuclear physics communities.

\vspace{2cm}

\begin{table}[ht!]
\centerline{
 \begin{tabular}{|l|l|l|l|l|l|l|l|l|l|l|}
  \hline\hline
  Decay & $T_{1/2}$ & $BR_{\nu}$ & $EC/{\beta^+}$ & $E_{GR}$ & $\Gamma_{GR}$ & $Q_{EC}$ & $E_{\nu}$ & $\Delta E_{\nu}$ \\ \hline
$^{148}Dy \rightarrow ^{148}Tb^{*}$ & $3.1 m$ & $1$ & $96/4$ & $620$ & $\approx0$ & $2682$ & $2062$ &  $\approx0$\\ \hline
$^{150}Dy \rightarrow ^{150}Tb^{*}$ & $7.2 m$ & $0.64$ & $100/0$ & $397$ & $\approx0 $ & $1794$ & $1397$ &  $\approx0$\\ \hline
$^{152}Tm2^{-} \rightarrow ^{152}Er^{*}$ & $8.0 s$ & $1$ & $45/55$ & $4300$ & $520$ & $8700$ & $4400$ & $520$ \\ \hline
$^{150}Ho2^{-} \rightarrow ^{150}Dy^{*}$ & $72 s$ & $1$ & $77/33$ & $4400$ & $400$ & $7400$ & $3000$ & $400$ \\ \hline
 \end{tabular}}
\caption{Four fast decays in the rare-earth region above $^{146}Gd$ leading to the
 giant Gamow-Teller resonance. Energies are given in keV. The first column gives the life-time, the second the branching ratio of the decay to neutrinos, the third the relative branching between electron capture and $\beta^+$, the fourth is the position of the giant GT resonance, the fifth its width, the sixth the total energy available in the decay, the seventh is the neutrino energy $E_\nu=Q_{EC}-E_{GR}$ and the eighth its uncertainty. }
\label{tab:1}
\end{table}

\section{Neutrino Flux}\label{sec5}
\noindent A neutrino (of energy $E_0$) that emerges from radioactive decay in an accelerator will
 be boosted in energy. At the experiment, the measured energy distribution as a function
 of angle $(\theta)$ and Lorentz gamma $(\gamma)$ of the ion at the moment of decay can 
be expressed as $E = E_0 / [\gamma(1- \beta \cos{\theta})]$. The angle $\theta$ in the 
formula expresses the deviation between the actual neutrino detection and the ideal detector
 position in the prolongation of one of the long straight sections of the Decay Ring of Figure \ref{betabeam}.
 The neutrinos are concentrated inside a narrow cone around the forward direction. If the ions
 are kept in the decay ring longer than the half-life, the energy distribution of the Neutrino 
Flux arriving to the detector in absence of neutrino oscillations is given by the Master Formula

\begin{equation}\label{master}
\frac{d^2N_\nu}{ dS dE}
= \frac{1}{\Gamma} \frac{d^2\Gamma_\nu}{dS dE} N_{ions}
\simeq \frac{\Gamma_\nu}{\Gamma} \frac{ N_{ions}}{\pi L^2} \gamma^2
\delta{\left(E - 2 \gamma E_0 \right)},
\end{equation}
with a dilation factor $\gamma >> 1$. It is remarkable that the result is given only in terms of the branching ratio 
and the neutrino energy and independent of nuclear models. In equation \ref{master}, $N_{ions}$ is the total number of 
ions decaying to neutrinos. For an optimum choice with $E \sim L$ around the
 first oscillation maximum, Eq. (\ref{master}) says that lower neutrino energies $E_0$ in the
 proper frame give higher neutrino fluxes.  The number of events will increase with higher
 neutrino energies as the cross section increases with energy. To conclude, in the forward 
direction the neutrino energy is fixed by the boost $E = 2 \gamma E_0$, with the entire neutrino 
flux concentrated at this energy. As a result, such a facility will measure the neutrino 
oscillation parameters by changing the $\gamma$'s of the decay ring (energy dependent measurement)
 and there is no need of energy reconstruction in the detector.

\section{Some Results }
\begin{figure}[ht!]
\centering
\includegraphics[width=13cm,height=10cm]{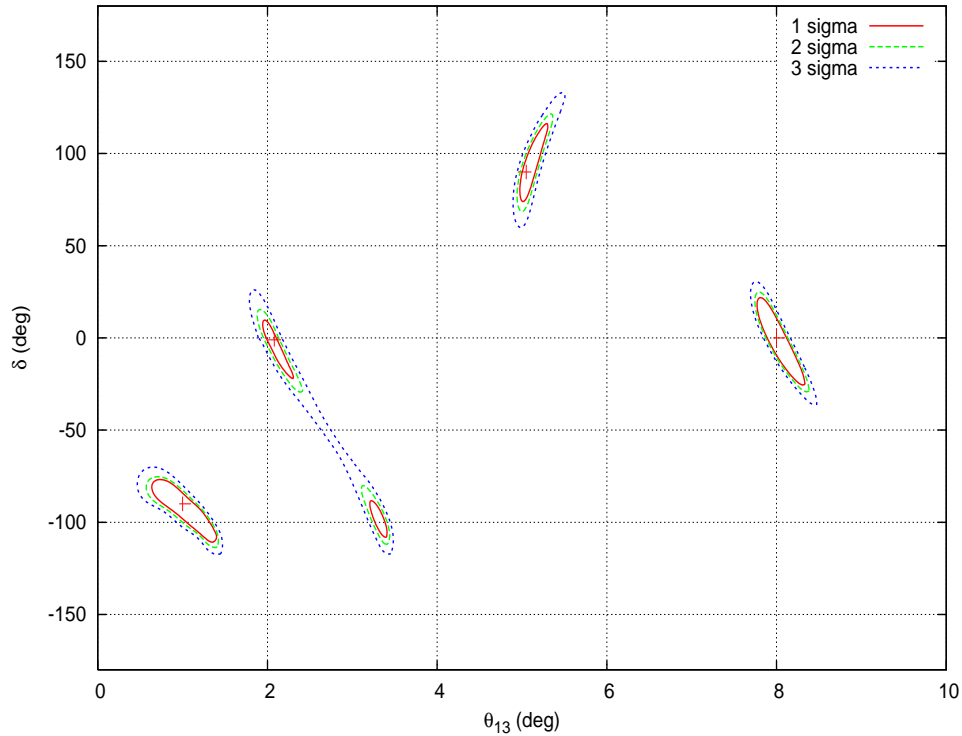}
\caption{Physics Reach for the presently unknown $(\theta_{13},\delta)$ parameters,
 using two definite energies in the electron-capture facility discussed in this paper.}
\label{fits}
\end{figure}

\noindent We have made a simulation study in order to reach conclusions about the measurability
of the unknown oscillation parameters. The ion type chosen is $^{150}Dy$, with neutrino energy at
rest given by $1.4$~MeV due to a unique nuclear transition from $100 \%$ electron capture in
going to neutrinos. Some $64 \%$ of the decay will happen as electron-capture, the rest goes 
through alpha decay. We have assumed that a flux of $10^{18}/y$ neutrinos at the end of the
 long straight section of the storage ring can be obtained (e.g. at the future European 
nuclear physics facility, EURISOL). We have taken two energies, defined by $\gamma_{max} = 195$
 as the maximum energy possible at CERN with the present accelerator complex, and a minimum,
$\gamma_{min} = 90$, in order to avoid atmospheric neutrino background in the detector below a certain energy. For
 the distance between source and detector we have chosen  $L = 130$ $km$ which equals the distance
 from CERN to the underground laboratory LSM in Frejus. The two values of $\gamma$ are
 represented as vertical lines in Fig.\ref{proba}. The detector has an active mass of $440$ $kton$ and
the statistics is accumulated during $10$ years, shared between the two runs at 
different $\gamma$'s, by detecting both appearance $(\nu_e \rightarrow \nu_{\mu})$ and
 disappearance  $(\nu_e \rightarrow \nu_e)$ events. Although the survival probability 
does not contain any information on the $CP$ phase, its measurement  helps in the cut of
 the allowed parameter region. The systematics will affect this cut, but one can expect a smaller 
 level of systematic error than in conventional neutrino beams or beta-beams, due to the precise 
 knowledge of the event energy. This is a subject for further exploration. The Physics 
Reach is represented by means of the plot in the parameters $(\theta_{13}, \delta)$ as
 given in Fig.\ref{fits}, with the expected results shown as confidence level lines for the
assumed values $(8^{\circ}, 0^{\circ})$, $(5^{\circ}, 90^{\circ})$, $(2^{\circ}, 0^{\circ})$ 
and $(1^{\circ}, -90^{\circ})$. The improvement over the standard beta-beam reach is due 
\ref{sec2}to the judicious choice of the energies to which the intensity is concentrated (see Fig.\ref{proba}):
whereas $\gamma=195$ leads to an energy above the oscillation peak with almost no dependence of
the $\delta-$phase, the value $\gamma=90$, leading to energies between the peak and the node, 
is highly sensitive to the phase of the interference. These two energies are thus complementary 
to fix the values of $(\theta_{13},\delta)$. The exclusion plot for the $\theta_{13}\neq 0$ sensitivity is particularly impressive \cite{Bernabeu:2005zs}. It is significant even at $1^{\circ}$, for all possible values of $\delta$.

To achieve a precise value of $\delta$, it has been shown \cite{Bernabeu:2005zs} that a combination of electron-capture neutrino beams with $\beta^{-}(^6He)$ antineutrino beams is very promising. An alternative could be to move to an smaller ratio $E/L$, where the sensitivity to $\delta$ is higher, by fixing the energy and going to the longer baseline of $650$~Km of Canfranc.

\section{Conclusions and Prospects}
\noindent The main conclusion is that {\bf the principle of an energy dependent
 measurement is working and a window is open to the discovery of CP violation in
 neutrino oscillations}, in spite of running at two energies only. The opportunity is
 better for higher values of the mixing angle $\theta_{13}$, the angle linked to the
 mixing matrix element $\vert U_{e3}\vert$ and for small mixing one would need to enter
 into the interference region of the neutrino oscillation by going to higher distance 
between source and detectors. To prove that the phase shift induced by $\delta$ in our EC design is due to a genuine CP-violating effect, one could combine in the facility the running with EC $^{150}Dy$ neutrinos and with $\beta^{-}$ $(^6He)$ antineutrinos.

The electron-capture facility will require a different approach to acceleration and storage of the ion beam compared to the standard beta-beam \cite{autin},
 as the ions cannot be fully stripped. Partly charged ions have a short vacuum life-time
\cite{franzke} due to a large cross-section for stripping through collisions with rest
 gas molecules in the accelerators. The isotopes discussed here have a half-life comparable
 to, or smaller than, the typical vacuum half-life of partly charged ions in an accelerator 
with very good vacuum. The fact that the total half-life is not dominated by vacuum losses
 will permit an important fraction of the stored ions sufficient time to decay through 
electron-capture before being lost out of the storage ring through stripping. A detailed study
 of production cross-sections, target and ion source designs, ion cooling and accumulation 
schemes, possible vacuum improvements and stacking schemes is required in order to reach a
 definite answer on the achievable flux. The experience gained at present on the acceleration and storage of partly charged ions and the calculations for the decay ring yield less than $5\%$ of stripping losses per minute. A Dy atom with only one electron left would still yield more than $40\%$ of the yield of the neutral Dy atom. 

With an isotope having an EC half-life of $1$ minute and a source rate of $10^{13}$ ions per second, a rate of $10^{18}$~$\nu$'s per year along one of the straight sections of the storage ring could be achieved \cite{priv-lind}. A new effort is on its way to revisit the rare-earth region on the nuclear cart and measure the EC properties of possible candidates.
The discovery of isotopes with half-lives of less than $1$ minute decaying mainly through electron-capture to a single Gamow-Teller resonance
 in super allowed transitions, certainly would push the concept of a monochromatic neutrino
 beam facility. The Physics Reach that we have explored here is 
impressive and demands such a study. Most important is a realistic simulation to determine the full capability of this facility.

\section{Acknowledgements}
\noindent The authors would like to thank our colleagues J.~Burguet-Castell and M.~Lindroos for an enjoyable collaboration. This paper has been supported by Spanish MEC and European FEDER under Grant FPA2005-01678.

\end{document}